\documentclass[10pt,twocolumn, amssymb, amsmath, aps, showpacs, footinbib, prb]{revtex4-1}
\usepackage{graphicx}

\newcommand{\eff}{\text{eff}}

\newcommand{\AFM}{\text{AFM}}
\newcommand{\FM}{\text{FM}}

\begin{document}

\title{$\beta$-Cu$_2$V$_2$O$_7$: a spin-$\frac12$ honeycomb lattice system}

\author{Alexander A. Tsirlin}
\email{altsirlin@gmail.com}
\author{Oleg Janson}
\email{janson@cpfs.mpg.de}
\author{Helge Rosner}
\email{Helge.Rosner@cpfs.mpg.de}
\affiliation{Max Planck Institute for Chemical Physics of Solids, N\"{o}thnitzer
Str. 40, 01187 Dresden, Germany}


\begin{abstract}
We report on band structure calculations and a microscopic model of the low-dimensional magnet $\beta$-Cu$_2$V$_2$O$_7$. Magnetic properties of this compound can be described by a spin-$\frac12$ anisotropic honeycomb lattice model with the averaged coupling $\bar J_1=60-66$~K. The low symmetry of the crystal structure leads to two inequivalent couplings $J_1$ and $J_1'$, but this weak spatial anisotropy does not affect the essential physics of the honeycomb spin lattice. The structural realization of the honeycomb lattice is highly non-trivial: the leading interactions $J_1$ and $J_1'$ run via double bridges of VO$_4$ tetrahedra between spatially separated Cu atoms, while the interactions between structural nearest neighbors are negligible. The non-negligible inter-plane coupling $J_{\perp}\simeq 15$~K gives rise to the long-range magnetic ordering at $T_N\simeq 26$~K. Our model simulations improve the fit of the magnetic susceptibility data, compared to the previously assumed spin-chain models. Additionally, the simulated ordering temperature of 27~K is in remarkable agreement with the experiment. Our study evaluates $\beta$-Cu$_2$V$_2$O$_7$ as the best available experimental realization of the spin-$\frac12$ Heisenberg model on the honeycomb lattice. We also provide an instructive comparison of different band structure codes and computational approaches to the evaluation of exchange couplings in magnetic insulators.
\end{abstract}

\maketitle

\section{Introduction}
Quantum magnetism is one of the most exciting and promising fields for exploring exotic ground states and unusual low-temperature properties. A variety of models and lattices lead to different regimes, ranging from simple, collinear long-range order to intricate and essentially quantum ground states.\cite{balents2010} One of the relevant examples is the honeycomb (hexagonal) lattice that comprises a two-dimensional (2D) network of regular hexagons. The spin-$\frac12$ Heisenberg model on the honeycomb lattice shows strong quantum fluctuations due to the low coordination number of 3.\cite{reger1989,oitmaa1992,richter} A weak interlayer coupling stabilizes the N\'eel ordering with the reduced sublattice magnetization of 0.54~$\mu_B$ (compare to 1~$\mu_B$ for a classical system).\cite{richter,castro2006,loew2009} Modifications of the model dramatically change its properties. For example, frustrating next-nearest-neighbor couplings induce a spin-liquid or a valence-bond-solid ground state.\cite{fouet2001,takano2006,mulder2010} The exactly solvable Kitaev model sets a specific arrangement of Ising-type interactions on the nearest-neighbor bonds of the honeycomb lattice and also leads to a spin-liquid ground state\cite{kitaev} that even sustains a partial disorder.\cite{willans} Despite the diversity of intriguing theoretical predictions, experimental studies remain scarce due to the lack of appropriate model materials.

Recent studies proposed several high-spin honeycomb lattice compounds, based on Ni$^{+2}$ (Ref.~\onlinecite{bani2v2o8}) and Mn$^{+4}$ (Ref.~\onlinecite{smirnova2009}). The iridates M$_2$IrO$_3$ (M = Li, Na) bear spin-$\frac12$ and likely show anisotropic exchange interactions on the honeycomb lattice.\cite{iridates} In contrast, the isotropic interaction regime is typically found in Cu$^{+2}$ or V$^{+4}$ compounds, well to be accounted by a Heisenberg model. The apparently hexagonal arrangement of Cu atoms in Na$_3$Cu$_2$SbO$_6$ is sometimes taken as an evidence of the honeycomb-lattice magnetism.\cite{na3cu2sbo6} However, the structural distortion and the orbital state of Cu make different bonds of the lattice inequivalent and induce one-dimensional magnetic behavior.\cite{na3cu2sbo6-INS} A more appropriate system could be InCu$_{2/3}$V$_{1/3}$O$_3$ but it shows intrinsic disorder due to the intermixing of spin-$\frac12$ Cu$^{+2}$ and non-magnetic V$^{+5}$ cations. Although the Cu atoms are believed to form small magnetic clusters on a honeycomb lattice,\cite{moeller2008,yehia2010} the influence of the non-magnetic part of this lattice is unclear and impedes the decisive comparison between experimental results and theoretical predictions. Thus, structurally ordered and well-characterized spin-$\frac12$ honeycomb-lattice materials with essentially isotropic, Heisenberg exchange are still lacking.

In low-dimensional magnets, the spatial arrangement of magnetic atoms is often deceptive, since the magnetic couplings do not show a clear correlation with distances between these atoms. In contrast, symmetries and overlaps of individual atomic orbitals play a crucial role.\cite{garrett1997,cucl,cuncn,schmitt2010} In the following, we provide an instructive example that demonstrates the relevance of these characteristics for the microscopic magnetic model. We show that the $\beta$-modification of Cu$_2$V$_2$O$_7$, previously considered as a spin-chain or a spin-dimer compound,\cite{pommer2003,ueda2008,yashima2009} is the best available realization of the spin-$\frac12$ Heisenberg model on the honeycomb lattice. To arrive at this unexpected conclusion, we first examine the crystal structure and analyze available experimental results (Sec.~\ref{intro}). Based on qualitative arguments, we rule out the previously assumed spin-chain scenario and further evaluate the relevant exchange couplings along with the appropriate spin model in a microscopic study (Sec.~\ref{band}). In Sec.~\ref{simul}, we perform simulations for the proposed spin model and confirm our results by a direct comparison to the experimental data. We conclude our work with Sec.~\ref{discussion} that reviews the unusual implementation of the honeycomb spin lattice in $\beta$-Cu$_2$V$_2$O$_7$ and provides an outlook for future experiments.
\begin{figure*}
\includegraphics{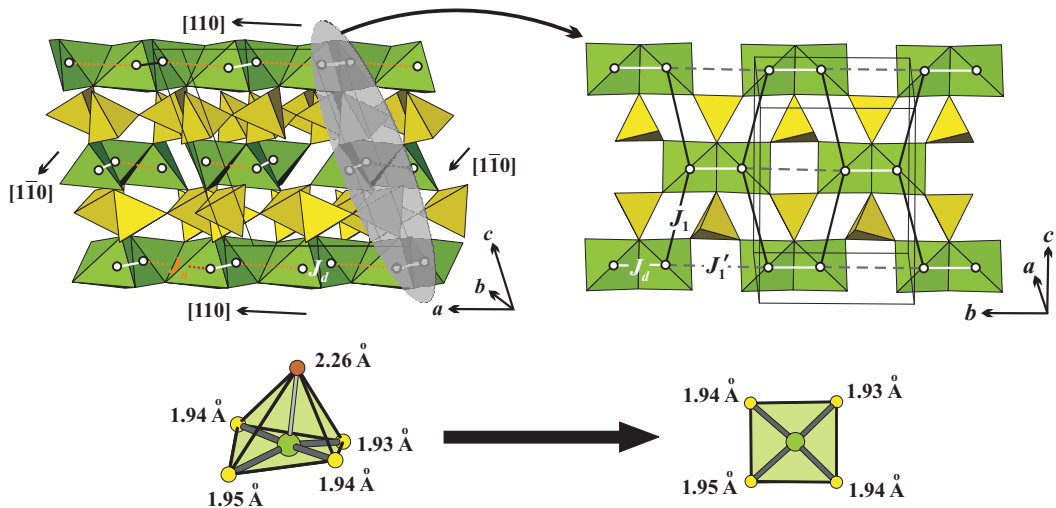}
\caption{\label{structure}
(Color online) Crystal structure of $\beta$-Cu$_2$V$_2$O$_7$ with copper and vanadium polyhedra colored in green (dark) and yellow (light), respectively. The left panel shows chains of edge-sharing CuO$_5$ pyramids running along $[110]$ and $[1\bar 10]$. The right panel presents a single layer of CuO$_4$ plaquettes in the $bc$ plane, also indicated by a shaded area in the left panel. Individual exchange couplings are labeled by dark solid ($J_1$), dashed ($J_1'$), light solid ($J_d$), and dotted ($J_a$) lines. The bottom part of the figure presents different representations of the Cu local environment: the CuO$_5$ pyramid (left) and the CuO$_4$ distorted plaquette (right). The spin lattice is shown in Fig.~\ref{honeycomb}.
}
\end{figure*}
\section{Crystal structure and magnetic properties}
\label{intro}
Experimental studies of $\beta$-Cu$_2$V$_2$O$_7$ reveal low-dimensional magnetic behavior with predominantly antiferromagnetic (AFM) couplings (the Curie-Weiss temperature $\theta\simeq 82$~K).\cite{ueda2008} The temperature dependence of the magnetic susceptibility has a broad maximum at 50~K, followed by the magnetic ordering transition at $T_N=26$~K.\cite{ueda2008} The compound is controversially ascribed to different spin models. Two experimental studies applied the uniform-chain description,\cite{pommer2003,ueda2008} whereas Touaiher \textit{et al}.\cite{touaiher2004} claimed a better susceptibility fit presuming the spin-dimer model. Further on, band structure calculations suggested two alternating couplings along the spin chains.\cite{yashima2009} Surprisingly, all the three proposals are quite different from the actual spin lattice. Below we demonstrate that empirical considerations in the experimental studies\cite{pommer2003,ueda2008,touaiher2004} do not take account of the peculiar electronic structure of $\beta$-Cu$_2$V$_2$O$_7$, while the computational study\cite{yashima2009} failed to resolve all the relevant exchange interactions in the compound. 

The crystal structure of $\beta$-Cu$_2$V$_2$O$_7$ is framework-like. Cu atoms have five-fold square-pyramidal coordination which is better described as "4+1" (see the bottom panel of Fig.~\ref{structure}). Four oxygen atoms in the $bc$ plane reveal shorter bonds to Cu ($1.93-1.95$~\r A)\cite{note1} and form a CuO$_4$ plaquette, typical for Cu$^{+2}$. The fifth oxygen atom shows a longer bond of 2.26~\r A which is roughly perpendicular to the plaquette along the $a$ axis. The distance between the apical oxygen atom and the mean plane of the CuO$_4$ plaquette is about 2.2~\r A. Considering the Cu polyhedron as a pyramid, one finds edge-sharing connections to one neighbor along the $b$ direction and to one neighbor along $a$, thus chains along $[110]$ and $[1\bar 10]$ are formed (upper left panel of Fig.~\ref{structure}). As the representation of the Cu environment is reduced to the plaquette (right part of Fig.~\ref{structure}), the connection along $a$ is broken, and we find Cu$_2$O$_6$ dimers of edge-sharing plaquettes (upper right panel of Fig.~\ref{structure}). Single VO$_4$ tetrahedra link such dimers within the $bc$ planes, while the pyrovanadate [V$_2$O$_7$] groups join the resulting layers into a framework. 

Overall, the structure looks quite complex. A first guess on the possible magnetic interactions is to consider shortest Cu--Cu distances for the edge-sharing pyramids: 2.95~\r A along $b$ (the interaction $J_d$ within the structural dimers) and 3.26~\r A along $a$ (the interaction $J_a$). The resulting bonds form an alternating $J_a-J_d$ chain proposed in Ref.~\onlinecite{yashima2009} (left panel of Fig.~\ref{structure}). Other models assume the dimer limit ($J_d\gg J_a$)\cite{touaiher2004} or the uniform chain limit $J_d\simeq J_a$.\cite{pommer2003,ueda2008} However, neither of these assumptions is correct due to the inappropriate choice of the leading couplings. The formation of a CuO$_4$ plaquette is a clear signature of the magnetic (half-filled) orbital of $d_{x^2-y^2}$ symmetry, typical for Cu$^{+2}$ with its $3d^9$ electronic configuration. In $\beta$-Cu$_2$V$_2$O$_7$, the plaquettes are slightly distorted. However, such a distortion has little effect on the crystal-field splitting, as confirmed by band structure calculations for $\beta$-Cu$_2$V$_2$O$_7$ (Sec.~\ref{band}) and for other Cu$^{+2}$ compounds.\cite{schmitt2010}

The magnetic orbital lies in the plane of the plaquette ($bc$) and does not overlap with the orbitals of the fifth, apical oxygen atom. This feature, further confirmed by band structure calculations (Fig.~\ref{wannier}), allows us to reduce the local environment of Cu to the CuO$_4$ plaquette, despite the presence of five Cu--O bonds in the crystal structure. The position of the magnetic orbital makes the coupling $J_a$ negligible (see Ref.~\onlinecite{nacu2o2} for a similar example). The coupling $J_d$ is still allowed for the $d_{x^2-y^2}$ orbital. However, this coupling corresponds to the Cu--O--Cu angle of $98.7^{\circ}$ implying sizable AFM and ferromagnetic (FM) contributions that might cancel each other.\cite{schmitt2010} These semi-empirical arguments are readily confirmed by our band structure calculations (Sec.~\ref{band}) and hint at relevant exchange couplings beyond the structural nearest neighbors.

After resorting to the CuO$_4$ plaquette description and considering the $bc$ plane as a potential magnetic layer, we find a remarkable similarity to the (VO)$_2$P$_2$O$_7$ structure (upper right panel of Fig.~\ref{structure}).\cite{note2} This similarity is not surprising due to the similar (effective) A$_2$B$_2$X$_7$ composition. In fact, $\beta$-Cu$_2$V$_2$O$_7$ and (VO)$_2$P$_2$O$_7$ also share similar interpretation problems, because the symmetry of the magnetic orbital determines possible superexchange pathways and renders some of the short V--V contacts magnetically ``inactive''. Thus, the early misleading conjecture\cite{johnston1987} on the spin ladder physics in (VO)$_2$P$_2$O$_7$ was based on the incorrect assumption of the strong coupling via the V--O--V superexchange pathway, although this pathway lacks any magnetically active orbital (similar to $J_a$ in $\beta$-Cu$_2$V$_2$O$_7$).\cite{garrett1997} As one further employs the structural analogy between $\beta$-Cu$_2$V$_2$O$_7$ and (VO)$_2$P$_2$O$_7$, a strong AFM coupling $J_1'$ and alternating $J_1'-J_d$ chains along the $b$ direction could be expected (right panel of Fig.~\ref{structure}). Yet, certain features of the electronic structure are different, thus even the $J_1'-J_d$ scenario of (VO)$_2$P$_2$O$_7$ has to be modified. Below, we show that the rearrangement of the cations (vanadium occupies the tetrahedral position and becomes non-magnetic, copper takes the vanadium position) strongly affects the electronic structure. As a result, the spin system becomes 2D and attains the honeycomb-lattice geometry.
\section{Methods}
\label{method}
To obtain a reliable microscopic model of $\beta$-Cu$_2$V$_2$O$_7$, we perform extensive density functional theory (DFT) band structure calculations using the full-potential local-orbital scheme (FPLO9.00-33).\cite{fplo} We apply the local density approximation (LDA) with the exchange-correlation potential by Perdew and Wang.\cite{perdew-wang} To cross-check the results, we used three alternative computational schemes: i) the generalized gradient approximation (GGA)\cite{pbe} for the exchange-correlation potential within FPLO; ii) the Vienna ab initio simulation package (VASP)\cite{vasp1,*vasp2} with the basis set of projected augmented waves;\cite{paw1,*paw2} iii) the tight-binding scheme for linearized muffin-tin orbitals in atomic spheres approximation (TB-LMTO-ASA),\cite{lmto} where a different procedure for the evaluation of the exchange couplings is implemented. The typical $k$ mesh included 1098 points in the symmetry-irreducible part of the first Brillouin zone for the crystallographic unit cell (LDA calculation) and 288~points for the supercell (DFT+$U$ calculations). We use the structural data\cite{note1} from Ref.~\onlinecite{structure1973} as well as the data from Ref.~\onlinecite{structure1989}. The results for the two structures match very well, with the largest deviation below 10~\% for one of the leading exchange couplings.

The bare LDA approach does not provide a realistic description for the electronic ground state of transition metal compounds due to the underestimate of strong correlation effects in the $3d$ shell. Nevertheless, LDA results can be taken as a reliable input for the modeling. Following this idea, we extract the relevant LDA valence bands and fit them with a one-orbital tight-binding (TB) model using Wannier functions (WFs) centered on Cu sites.\cite{wannier} The application of the WF technique leads to the unambiguous evaluation of hopping parameters $t_i$ and provides a clear picture of the magnetic orbitals. Further on, we introduce the hopping parameters into a Hubbard model with the effective on-site Coulomb repulsion $U_{\eff}$. The Hubbard model is then reduced to the Heisenberg model to describe the low-lying excitations. This reduction is justified by the half-filling and by the strongly correlated ($t_i\ll U_{\eff}$) regime. The parameters of the Heisenberg model are expressed as $4t_i^2/U_{\eff}$ and describe the AFM part $J_i^{\AFM}$ of the exchange. The parameter $U_{\eff}$ is fixed at 4.5~eV, according to previous studies.\cite{cucl,schmitt2010,janson2007} 

The second approach to the evaluation of the exchange couplings includes a mean-field treatment of correlation effects via the local spin density approximation (LSDA)+$U$ (or the related GGA+$U$) method. Two options are possible: i) total energies for a set of collinear spin configurations are mapped onto a classical Heisenberg model (supercell approach); ii) the exchange integrals are treated as second derivatives of the energy with respect to the rotation of the magnetic moments; such derivatives are calculated via matrix elements of the Green's function [Lichtenstein exchange integral procedure (LEIP)].\cite{leip} The former approach is realized in FPLO and VASP, while the latter is implemented in the TB-LMTO-ASA code. The on-site exchange parameter of the LSDA+$U$ method was fixed at $J_{3d}=1$~eV. The choice of the Coulomb repulsion parameter $U_{3d}$ is further discussed in Sec.~\ref{lsda+u}.

The DFT-based microscopic model was challenged by the experimental magnetic susceptibility data and the magnetic ordering temperature, taken from Ref.~\onlinecite{ueda2008}. The respective quantities as well as the high-field magnetization curves, static structure factors, and spin correlations were computed via quantum Monte-Carlo (QMC) simulations with the \texttt{loop} algorithm\cite{looper} or the directed loop algorithim in the stochastic series expansion representation,\cite{dirloop} as implemented in the ALPS simulation package.\cite{alps} The simulations are done for finite lattices with periodic boundary conditions. The typical lattice size was $16\times 16$ (512~sites) for the 2D model (magnetic susceptibility, high-field magnetization, spin correlations, ordered moment) and up to $16\times 16\times 12$ (6144~sites) for the three-dimensional (3D) model (magnetic ordering temperature). The results obtained on lattices of different size ensured the lack of appreciable finite-size effects. To determine the ordered moment, we calculated static structure factors at a constant temperature $T/\bar J_1=0.01$ and performed finite-size scaling, following Ref.~\onlinecite{castro2006}.
\begin{figure}
\includegraphics{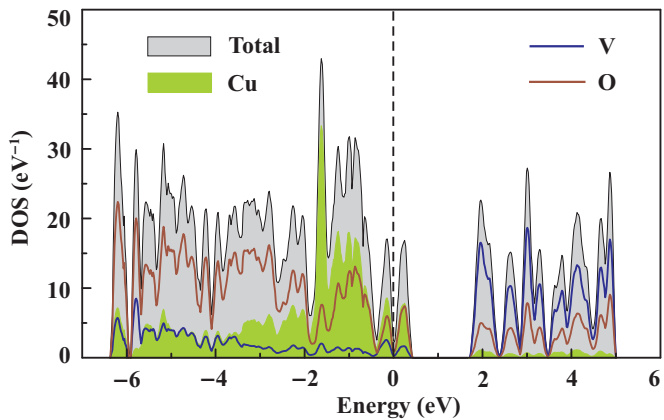}
\caption{\label{dos}
(Color online) Total and atomic-resolved LDA DOS for $\beta$-Cu$_2$V$_2$O$_7$. The Fermi level is at zero energy.
}
\end{figure}
\section{Evaluation of the exchange couplings}
\label{band}
\subsection{LDA-based model}
\label{lda}
\begin{figure}
\includegraphics{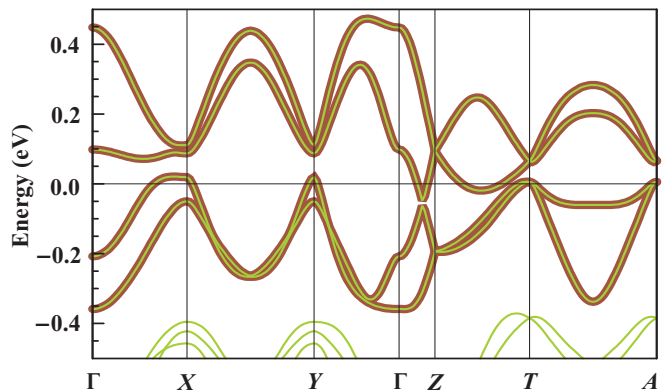}
\caption{\label{bands}
(Color online) LDA bands at the Fermi level (thin light lines) and the fit of the tight-binding model (thick dark lines). The notation of the $k$ points is as follows: $\Gamma(0,0,0)$, $X(0.5,0,0)$, $Y(0,0.5,0)$, $Z(0,0,0.5)$, $T(0.5,0,0.5)$, and $A(0,0.5,0.5)$, and the coordinates are given along $k_x, k_y$, and $k_z$ in units of the respective reciprocal lattice parameters.
}
\end{figure}

The main features of the LDA density of states (DOS, Fig.~\ref{dos}) are typical for Cu$^{+2}$ compounds.\cite{janson2007,schmitt2010,dioptase} Essentially, the valence bands are a mixture of Cu and O states with a small contribution of vanadium. These bands lie between $-6.5$~eV and $0.5$~eV and account for the bonding between Cu and the five neighboring oxygen atoms in the CuO$_5$ pyramid. The bands with the predominant vanadium contribution are found above 1.5~eV in agreement with the oxidation state of +5 for V atoms. The nearly isolated band complex at the Fermi level (Fig.~\ref{bands}) is mostly formed by the magnetic Cu $d_{x^2-y^2}$ orbital and the hybridizing $2p_{\sigma}$ orbitals of the four oxygen atoms comprising the CuO$_4$ plaquette. The magnetic orbital lies in the plane of the plaquette and determines possible superexchange pathways (see Sec.~\ref{intro}). The energy spectrum is metallic due to the well-known shortcoming of LDA for strongly correlated systems. LSDA+$U$ reproduces the insulating spectrum in agreement with the experimental identification of $\beta$-Cu$_2$V$_2$O$_7$ as a magnetic insulator.\cite{pommer2003,ueda2008,optics}

\begin{table}
\caption{\label{tbm}
Cu--Cu distances (in~\r A), leading hopping parameters $t_i$ (in~meV) of the TB model, and the resulting AFM contributions to the exchange integrals $J_i^{\AFM}$ (in~K). The exchange integrals are derived as $4t_i^2/U_{\eff}$ with the effective on-site Coulomb repulsion $U_{\eff}=4.5$~eV.
}
\begin{ruledtabular}
\begin{tabular}{rcrr}
        &  Distance  &   $t_i$ &  $J_i^{\AFM}$ \\\hline
 $J_d$  &   2.95     &   148   &      227      \\
 $J_a$  &   3.26     &    36   &       13      \\
 $J_1$  &   5.18     &    97   &       96      \\
 $J_1'$ &   5.25     &  $-84$  &       73      \\
 $J_{\perp}$ & 7.32 &     35   &       13      \\
\end{tabular}
\end{ruledtabular}
\end{table}

The TB fit of the Cu $d_{x^2-y^2}$ bands (Fig.~\ref{bands}) yields three leading AFM interactions in the $bc$ plane: within the dimers ($J_d$), between the dimers along the $b$ direction ($J_1'$), and between the dimers along $[210]$ or $[\bar 210]$ ($J_1$), see Table~\ref{tbm}. The nearest-neighbor interaction along $a$ ($J_a$) is one of the leading magnetic couplings between the $bc$ planes, but it is much weaker than the next-nearest-neighbor in-plane couplings $J_1$ and $J_1'$. This supports our empirical, qualitative considerations in Sec.~\ref{intro}. Contrary to the spin-chain scenario, we propose a 2D model with leading exchange couplings in the $bc$ plane. At first glance, the model approach would support the spin-dimer interpretation of Ref.~\onlinecite{touaiher2004} with the largest coupling $J_d$. However, the nearly $90^{\circ}$ superexchange regime of $J_d$ implies a large FM contribution that strongly modifies the TB scenario (see Sec.~\ref{lsda+u}).\cite{mazurenko2007} 

The couplings between the $bc$ planes are below 15~K. In addition to the nearest-neighbor interaction $J_a$, we find the long-range couplings $J_{\perp}$ with similar energy. Despite the short Cu--Cu distance, $J_a$ is small, since the magnetic orbital of Cu does not overlap with the orbitals of the apical oxygen atom and makes the Cu--O--Cu superexchange impossible. The couplings $J_{\perp}$ run via the V$_2$O$_7$ groups along (approximately) $[111]$ and $[1\bar 11]$ directions. Further hoppings in the TB model are below 25~meV, i.e., the respective $J^{\AFM}$'s do not exceed $5-7$~K and can be neglected within the minimum microscopic model.

\begin{figure}
\includegraphics[scale=1]{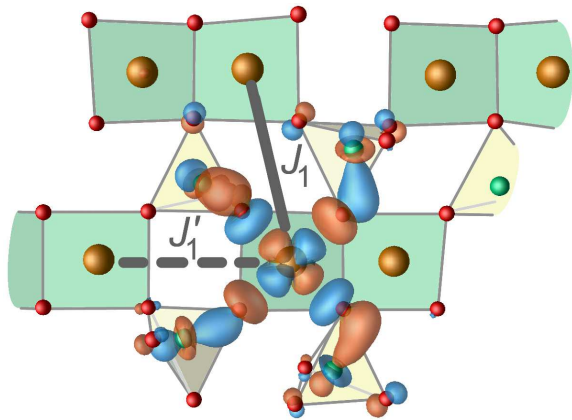}
\caption{\label{wannier}
(Color online) The Wannier function centered on the Cu site. 
}
\end{figure}

The calculated WFs (Fig.~\ref{wannier}) give a visual representation of the orbitals contributing to the superexchange. Each WF includes the Cu $d_{x^2-y^2}$ orbital and the $p_{\sigma}$ oxygen orbitals along with the smaller contribution of vanadium orbitals. Since all the orbitals lie in the $bc$ plane, the interplane couplings are relatively weak. Regarding the in-plane couplings, we note that oxygen contributions to WFs on the neighboring Cu sites show a nearly $90^{\circ}$ arrangement for $J_d$. A similar arrangement is found for vanadium contributions with respect to $J_1$ and $J_1'$. This causes a FM contribution to the exchange due to the Hund's coupling on the ligand site. 

Mazurenko \textit{et al}.\cite{mazurenko2007} developed a model for the ideal $90^{\circ}$ superexchange geometry and calculated the FM part of the exchange: $J^{\FM}=-2\beta^4J_HN_l$, where $\beta$ is the ligand contribution to the WF, $J_H$ is Hund's coupling on the ligand site, and $N_l$ is the number of ligands where the WFs overlap (see also Ref.~\onlinecite{cuncn}). In our case, $\beta^2_{\text{O}}=0.10-0.12$, $\beta^2_{\text{V}}\simeq 0.03$, and $N_l=2$. Assuming $J_H=1.5$~eV,\cite{mazurenko2007,cuncn} we find $J_d^{\FM}\simeq -850$~K and $J_1^{\FM}\simeq J_1'^{\FM}\simeq -60$~K. These values are rather overestimated (compare Tables~\ref{tbm} and~\ref{supercell}: $J_d^{\FM}\simeq -220$~K, $J_1^{\FM}\simeq -10$~K, and $J_1'^{\FM}\simeq -15$~K), yet they identify a sizable FM contribution to $J_d$ along with order of magnitude smaller FM contributions to $J_1$ and $J_1'$. The overestimate is likely caused by the complex geometry of the system: the Cu--ligand distances are inequivalent, whereas the angles exceed $90^{\circ}$ (e.g., the Cu--O--Cu angle for $J_d$ is $98.7^{\circ}$), thus significantly reducing the Hund's coupling.

\subsection{DFT+$U$}
\label{lsda+u}
\begin{figure}
\includegraphics{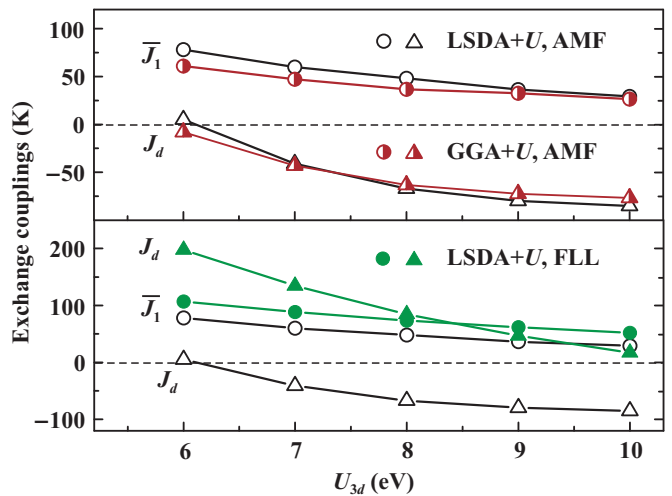}
\caption{\label{j-vs-u}
(Color online) Exchange couplings $\bar J_1=(2J_1+J_1')/3$ (circles) and $J_d$ (traingles) calculated with FPLO at different values of $U_{3d}$, the Coulomb repulsion parameter of DFT+$U$. The upper panel compares LSDA+$U$ and GGA+$U$ for the AMF double-counting correction scheme, while the bottom panel compares two double-counting corrections for the LSDA+$U$ functional (see text).
}
\end{figure}

To evaluate full exchange integrals via DFT+$U$ calculations, we consider the relevant in-plane couplings $J_d,J_1$, and $J_1'$, as well as the two interplane couplings $J_a$ and $J_{\perp}$. The results are collected in Fig.~\ref{j-vs-u} and Table~\ref{supercell}. Most of the couplings are weakly dependent on the computational procedure, but the estimate of $J_d$ is highly sensitive to the $U_d$ value and to details of the DFT+$U$ method. The investigation of the respective methodological problems constitutes the main part of this section.

We start with the FPLO results shown in Fig.~\ref{j-vs-u}. Different exchange-correlation potentials (LSDA+$U$ vs. GGA+$U$, upper panel) have little effect on the exchange couplings. In contrast, the double-counting correction (DCC) scheme of DFT+$U$ has a stronger effect on $J$'s and especially influences $J_d$ (bottom panel of the figure). The DFT+$U$ method adds an explicit mean-field energy term to account for the Coulomb repulsion in the correlated shell. Then, a DCC is required. The DCC term subtracts a part of the repulsion energy that is already contained in LSDA (GGA). Several DCC schemes are available, but their differences remain weakly explored.\cite{petukhov2003,ylvisaker} In the FPLO code, two DCC schemes are implemented: around-mean-field (AMF)\cite{AMF} and fully-localized-limit (FLL).\cite{FLL}

\begin{table}
\caption{\label{supercell}
Exchange couplings calculated within LSDA+$U$. The columns list individual exchange integrals (in~K), the Coulomb repulsion parameter $U_{3d}$ of LSDA+$U$ (in~eV), the band structure code, and the double-counting correction. The solutions consistent with the experiment are marked with the bold typeface.
}
\begin{ruledtabular}
\begin{tabular}{rrrrr@{\hspace{2em}}rrr}
  $J_d$ & $J_a$ & $J_1$ & $J_1'$ & $J_{\perp}$ & $U_{3d}$ & Code & DCC \\\hline\medskip
 \textbf{5}  &  $\mathbf{-3}$ &  \textbf{87} & \textbf{61} & \textbf{17} & \textbf{6} & \textbf{FPLO} & \textbf{AMF} \\
   198  & $-9$  &  100  &   121  &    37       &    6     & FPLO & FLL \\
   228  & $-10$ &  108  &   125  &    38       &    7     & VASP & FLL \\\medskip
   232  & $0$   &  83   &   50   &    14       &   9.5    & LMTO & FLL \\
 \textbf{17}  & $\mathbf{-3}$ & \textbf{48} & \textbf{60} & \textbf{18} & \textbf{10} & \textbf{FPLO} & \textbf{FLL} \\
\end{tabular}
\end{ruledtabular}
\end{table}
Band structure calculations for a family of copper oxides have rather firmly established the proper $U_{3d}$ value of $6-7$~eV to reproduce the experimental exchange couplings within AMF.\cite{janson2007,schmitt2010,li2cuo2,dioptase} Indeed, $U_{3d}=6$~eV in AMF leads to a reasonable scenario (first row of Table~\ref{supercell}). The interactions $J_1$, $J_1'$, and $J_{\perp}$ are nearly unchanged compared to the TB estimates in Table~\ref{tbm}. This implies weak FM contributions in agreement with the long-range nature of these couplings. In contrast, the nearest-neighbor coupling $J_d$ is reduced almost down to zero due to the superexchange pathway via the $98.7^{\circ}$ bond angle. The other nearest-neighbor coupling $J_a$ also has a FM contribution and becomes weakly FM. Since $|J_a|\leq 5$~K, this interaction can even be omitted in a minimum microscopic model. As the $U_{3d}$ value is increased above 6~eV, the interactions $J_1$ and $J_1'$ are slightly reduced, while $J_d$ becomes FM.

Switching to FLL, we find a different scenario (Table~\ref{supercell} and the bottom part of Fig.~\ref{j-vs-u}). The couplings $J_1$ and $J_1'$ are close to the AMF estimates. However, the short-range coupling $J_d$ is about 200~K at $U_{3d}=6$~eV. As $U_{3d}$ is increased to 10~eV, $J_d$ becomes relatively smaller, resembling the AMF result at $U_{3d}=6$~eV. Overall, we can reliably estimate $J_1, J_1'=50-100$~K, while $J_d$ can be either: i) weak AFM (AMF, $U_{3d}=6$~eV or FLL, $U_{3d}=10$~eV); ii) strong AFM (FLL, $U_{3d}=6$~eV); iii) strong FM (AMF, $U_{3d}\geq 8$~eV). In Sec.~\ref{simul}, we carefully analyze the experimental data that unambiguously favor the first scenario. Unfortunately, this choice can not be made on purely computational (i.e., \textit{ab initio}) grounds. The DFT+$U$ calculations are typically compared to the experimental band gap and the ordered magnetic moment (sublattice magnetization). The sublattice magnetization of $\beta$-Cu$_2$V$_2$O$_7$ has not been reported and, anyway, should be subject to quantum effects (Sec.~\ref{simul}) that lie beyond DFT+$U$ and preclude the reliable comparison. The experimental estimate of the band gap ($E_g=1.8$~eV\cite{optics}) is in good agreement to both AMF ($E_g=2.0$~eV) and FLL ($E_g=1.75$~eV) results at $U_{3d}=6$~eV. As $U_{3d}$ is increased to 10~eV, the band gap is somewhat overestimated ($E_g=2.3$~eV in FLL). Thus, the computational arguments would rather prefer a ``typical'' $U_{3d}$ of 6~eV to describe the correlated electronic system of $\beta$-Cu$_2$V$_2$O$_7$. Yet, neither AMF nor FLL can be chosen unambiguously at this point.

The DCC problem is not unique to the FPLO code, although in other codes it may be left out due to the lack of alternative DCC schemes implemented. In VASP, only the FLL and the related Dudarev's approach\cite{dudarev} to the DCC are available. According to previous studies,\cite{mentre2009,cucl2-kremer} we select $U_{3d}=7$~eV and find a remarkable agreement with the FPLO FLL results at $U_{3d}=6$~eV (see Table~\ref{supercell}). The 1~eV difference in $U_{3d}$ is a typical offset due to the different basis sets (the Coulomb repulsion potential is applied to the atomic $3d$ orbitals which are basis-dependent). Further on, we used the LMTO code and determined $U_{3d}$ from the constrained LDA procedure.\cite{constrained} The resulting $U_{3d}$ of 9.5~eV is in agreement with previous reports\cite{cuncn,mazurenko2007,mazurenko2008} and also leads to the typical FLL result with the large $J_d$.\cite{note7} Overall, different band structure codes converge to the same FLL solution for the exchange integrals. In this solution, $J_d$ largely exceeds $J_1$ and $J_1'$ and contradicts the experimental energy scale (see Sec.~\ref{simul}). To get a realistic solution with small $J_d$, one should either switch to the AMF DCC or take a larger $U_{3d}$ value of about 10~eV for FPLO (and even larger values for VASP and LMTO). 

The problem of the DCC choice is not specific to $\beta$-Cu$_2$V$_2$O$_7$, although in this compound it is probably most evident. Recently, we have shown that the AMF and FLL flavors of LSDA+$U$ lead to different results for several Cu$^{+2}$ compounds.\cite{cuncn,dioptase,bicu2po6} Similar to $\beta$-Cu$_2$V$_2$O$_7$, experimental data then favor the results at $U_{3d}=6-7$~eV for AMF and at $U_{3d}=9-10$~eV for FLL within FPLO.\cite{dioptase} This empirical recipe helps to adjust the $U_{3d}$ value/the DCC scheme and to obtain realistic estimates of the exchange couplings. Still, a more general and comprehensive theoretical study of this computational effect is highly desirable.

The methodological part of the DFT+$U$ study can be summarized as follows:

i) The DCC scheme of DFT+$U$ has influence on the exchange couplings and especially affects the short-range interactions;

ii) The difference between the AMF and FLL flavors of DFT+$U$ can be partly balanced by adjusting $U_{3d}$. However, the FLL results require unusually large $U_{3d}$ values that look unjustified with respect to the typical estimates (e.g., from constrained LDA or from the comparison to the experimental band gap).

\begin{figure}
\includegraphics{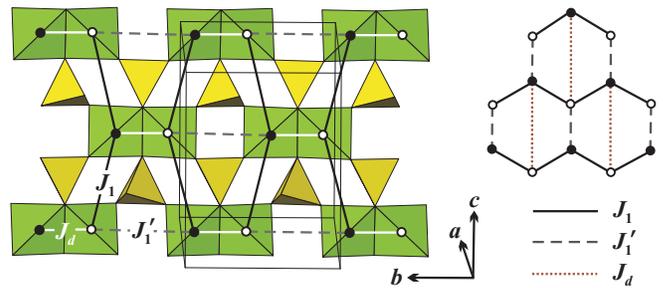}
\caption{\label{honeycomb}
(Color online) Crystal structure of $\beta$-Cu$_2$V$_2$O$_7$ and the sketch of the spin model with the leading couplings $J_1$ (solid lines) and $J_1'$ (dashed lines). Dotted lines show $J_d$, the third-neighbor coupling on the honeycomb lattice. Open and shaded circles denote Cu atoms with opposite spins in the magnetically ordered ground state.
}
\end{figure}
Switching back to the low-dimensional magnetism, we are able to construct a consistent microscopic picture, based on the TB analysis and the LSDA+$U$ calculations. We establish a 2D spin model for $\beta$-Cu$_2$V$_2$O$_7$. The leading AFM interactions are $J_1$ and $J_1'$ that act in the $bc$ plane. The non-frustrated interplane couplings $J_{\perp}$ amount to $10-15$~K. The nearest-neighbor coupling $J_d$ is somewhat ambiguous. However, the estimate of $J_d\simeq 200$~K clearly contradicts the experimental energy scale. A thorough analysis in Sec.~\ref{simul} shows that $J_d$ is essentially a weak coupling, well below $J_1$ and $J_1'$. The resulting $J_1-J_1'$ spin system reveals three bonds per site and represents the anisotropic honeycomb lattice (Fig.~\ref{honeycomb}). The coupling $J_d$ between the structural nearest neighbors connects third neighbors on the spin lattice. Despite their complex structural arrangement, $J_{\perp}$ interactions can be considered as uniform interplane couplings.
\section{Simulations and comparison to the experiment}
\label{simul}
\begin{figure}
\includegraphics{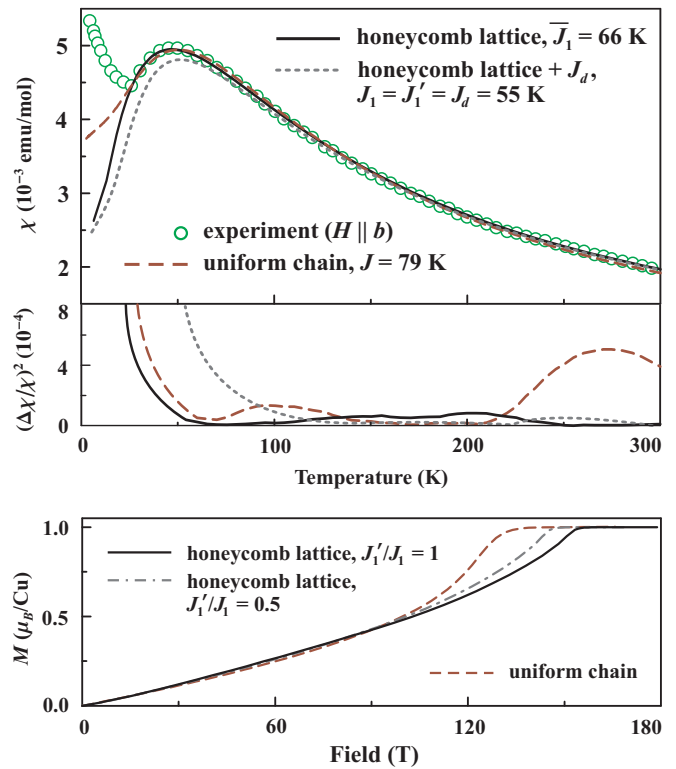}
\caption{\label{chi}
(Color online) Upper panel: magnetic susceptibility of $\beta$-Cu$_2$V$_2$O$_7$ and the fits of the honeycomb-lattice (solid line), the uniform-chain (long-dashed line), and the honeycomb lattice with AFM $J_d$ (short-dashed line) models. Experimental data are taken from Ref.~\onlinecite{ueda2008}. Difference curves are shown under the figure. Bottom panel: magnetization curves for the isotropic honeycomb lattice (solid line), anisotropic honeycomb lattice (dash-dotted line, $J_1'/J_1=0.5$), and the uniform chain (dashed line) calculated at $T/\bar J_1=0.03$.
}
\end{figure}

To challenge the proposed spin model by the experimental data and to investigate the properties of this model, we perform extensive QMC simulations. We start with the magnetic susceptibility\cite{note5} that was previously fitted with the expression for the uniform chain model.\cite{ueda2008} This fit yields $J\simeq 79$~K and shows slight deviations from the experiment both at high temperatures and near the maximum (Fig.~\ref{chi}). The model of the isotropic honeycomb lattice with $\bar J_1=66$~K produces a significantly better fit which is in remarkable agreement with the experiment down to $T_N$ (see the difference plot in Fig.~\ref{chi}). Below $T_N$, the susceptibility is strongly dependent on the direction of the applied field and can not be reproduced within the Heisenberg model that assumes isotropic exchange.

The low symmetry of the $\beta$-Cu$_2$V$_2$O$_7$ structure leads to a weakly anisotropic honeycomb lattice with inequivalent couplings $J_1$ and $J_1'$. However, this spatial anisotropy does not affect the magnetic susceptibility. We were able to fit the data down to $T_N$ with $\bar J_1=(2J_1+J_1')/3=60-66$~K for $J_1'/J_1=0.5-1.3$ (the averaging includes two $J_1$ and one $J_1'$, because there are two $J_1$ bonds and one $J_1'$ bond per site). This situation resembles the anisotropic frustrated square lattice\cite{tsirlin2009} where the moderate spatial anisotropy has little effect on the magnetic susceptibility. The fitted $g$ value ($1.9-2.2$ depending on the field direction, in reasonable agreement with the powder-averaged $g\simeq 2.1$ from electron-spin-resonance measurements\cite{pommer2003}) weakly depends on $J_1'/J_1$ and can not be used to evaluate the degree of spatial anisotropy. Irrespective of the precise $J_1'/J_1$ ratio, the averaged coupling $\bar J_1$ is in excellent agreement with our computational estimates of $50-100$~K for $J_1$ and $J_1'$ (see Tables~\ref{tbm} and~\ref{supercell}). 

\begin{figure}
\includegraphics{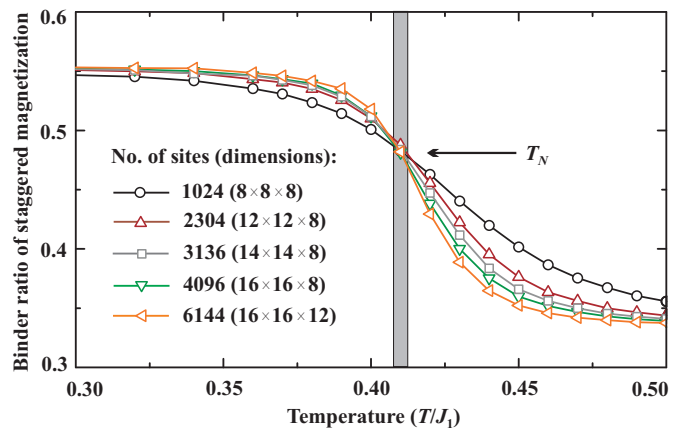}
\caption{\label{fig_binder}
(Color online) Binder ratio of the staggered magnetization $\langle m_s^4\rangle/\langle m_s^2\rangle^2$ for the isotropic honeycomb lattice with the interlayer coupling $J_{\perp}/\bar J_1=0.2$. Different symbols correspond to finite lattices of different size (note that the unit cell comprises two sites). The shaded bar shows the N\'eel temperature $T_N/\bar J_1\simeq 0.41$.
}
\end{figure}
To resolve the ambiguity regarding the size of $J_d$, we extended the model and introduced the third-neighbor coupling on the honeycomb lattice (right panel of Fig.~\ref{honeycomb}). A small AFM $J_d$ ($J_d/\bar J_1=0.2$) does not lead to any visible changes in the fit and slightly renormalizes $\bar J_1$. The effect of the moderate interplane coupling $J_{\perp}=0.2\bar J_1$ is similar to that of $J_d$. However, an AFM intradimer coupling on the order of $\bar J_1$ ($J_d=\bar J_1$) notably reduces the susceptibility maximum and leads to a poor fit of the data below 80~K (Fig.~\ref{chi}). A larger $J_d$ of $150-200$~K (as suggested by the FLL calculations, see Table~\ref{supercell}) clearly contradicts the experimental energy scale. Finally, the sizable FM coupling $J_d$ would strongly frustrate the spin lattice and impede the long-range ordering. Our simulations for the unfrustrated lattice show excellent agreement with the experimental ordering temperature (see below), hence the scenario of strong FM $J_d$ is unlikely. We conclude that the coupling $J_d$ is weak, while the leading couplings are $J_1$ and $J_1'$ forming an anisotropic honeycomb lattice. 

The magnetic ordering temperature $T_N$ enables further comparison to the experimental data. To achieve the long-range ordering, we switch to a 3D model with an interplane coupling $J_{\perp}\simeq 13$~K $\simeq 0.2\bar J_1$ (cf. Table~\ref{supercell}). In a 3D model, the ordering transition is evidenced by a kink in the susceptibility curve or by a sharp increase in the staggered magnetization. To get a more accurate estimate, we calculate the Binder ratio $\langle m_s^4\rangle/\langle m_s^2\rangle^2$ where $m_s$ is the staggered magnetization. The Binder ratio demonstrates a significant change upon the transition (Fig.~\ref{fig_binder}). The intersection of the curves obtained for finite lattices of different size can be taken as a numerical estimate of $T_N$.\cite{binder} We find $T_N=0.41\bar J_1\simeq 27$~K in remarkable agreement with the experimentally observed $T_N=26$~K.\cite{ueda2008} This comparison further supports our spin model. Similar to the magnetic susceptibility, the moderate spatial anisotropy ($J_1'/J_1=0.5-1.3$) has little effect on $T_N$ (at $J_1'/J_1=0.5$, $T_N/\bar J_1=0.37$; at $J_1'/J_1=1.3$, $T_N/\bar J_1=0.40$). 

The temperature dependence of the magnetic susceptibility favors the honeycomb-lattice description of $\beta$-Cu$_2$V$_2$O$_7$ and yields the averaged coupling $\bar J_1$. However, the magnitude of the spatial anisotropy $J_1'/J_1$ can not be precisely estimated. For the anisotropic frustrated square lattice compounds, high-field magnetization measurements are the most suitable tool to evaluate the spatial anisotropy.\cite{high-field} Therefore, we also simulated magnetization curves of $\beta$-Cu$_2$V$_2$O$_7$ (see the lower panel of Fig.~\ref{chi}). The saturation field is slightly reduced for $J_1'/J_1<1$ due to the decrease in $\bar J_1$. The uniform-chain scenario would lead to an even lower saturation field of 130~T, compared to $145-155$~T for the honeycomb lattice. However, the easily accessible field region up to 60~T shows negligible differences between the simulated curves. We conclude that high-field magnetization measurements could be a helpful tool to study the system, though only once the region around the saturation field ($130-155$~T) is reached.

\begin{figure}
\includegraphics[width=8.6cm]{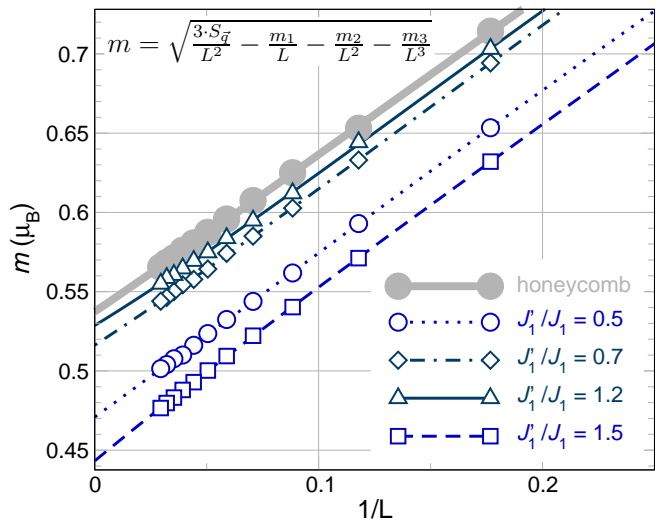}
\caption{\label{fig_moment}
(Color online) Finite-size scaling for the ordered moment (sublattice magnetization, $m$) of the anisotropic honeycomb lattice model. $L$ is the period of the finite lattice (cluster size) and $S_{\mathbf{\vec q}}$ is the static structure factor at the propagation vector $\mathbf{\vec q}$.
}
\end{figure}
Since the honeycomb spin lattice in $\beta$-Cu$_2$V$_2$O$_7$ may be slightly anisotropic ($0.5\leq J_1'/J_1\leq1.3$), we study the influence of this spatial anisotropy on the ground state. The isotropic ($J_1'/J_1=1$) lattice leads to the largest $m$ of about 0.54~$\mu_B$. The spatial anisotropy reduces the dimensionality of the system and, consequently, reduces $m$ down to 0.47~$\mu_B$ for $J_1'/J_1=0.5$ and down to 0.44~$\mu_B$ for $J_1'/J_1=1.5$ (Fig.~\ref{fig_moment}). Although the magnitude of the effect is $15-20$~\%, it may be hard to resolve experimentally due to the low absolute values of $m$ and the high uncertainty of experimental estimates (typically, above 10~\%). We also note that the interlayer coupling will rather increase the dimensionality leading to a slight increase in the $m$ value.

\begin{figure}
\includegraphics[width=8.6cm]{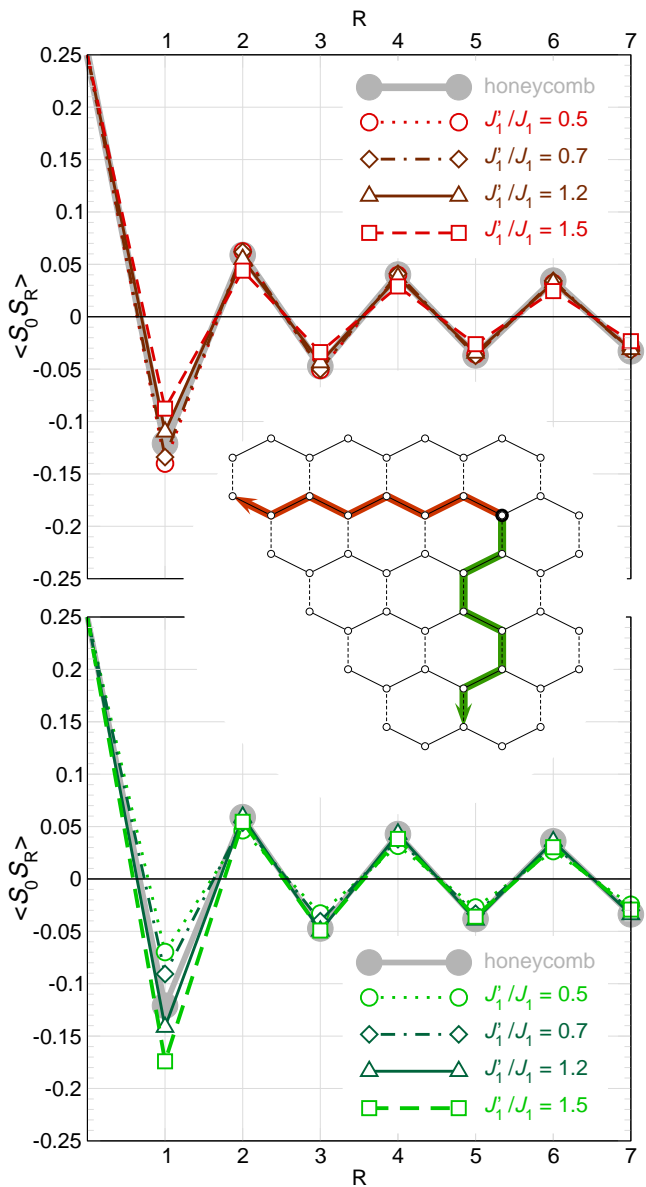}
\caption{\label{cu2v2o7_fig_correl}
(Color online) Spin correlations for different spatial anisotropies $J_1'/J_1$ compared to the isotropic honeycomb lattice (bold lines) along the $J_1$ chains (top) and in the perpendicular direction (bottom). The inset shows the respective paths; $J_1$ and $J_1'$ are depicted by solid and
dashed lines, respectively.
}
\end{figure}

Spin correlations are a sensitive probe for the ground-state properties. Therefore, we calculate the expectation values ${\langle}S_0S_{\text{R}}{\rangle}$ for $\text{R}=1-7$ running along the
$J_1$ chains and in the perpendicular direction (see the inset of Fig.~\ref{cu2v2o7_fig_correl}). For the isotropic (ideal) honeycomb lattice, both paths yield the same correlations (Fig.~\ref{cu2v2o7_fig_correl}, bold lines). The spatial anisotropy $J_1'/J_1\neq1$ slightly affects the correlations: for $J_1'/J_1<1$, the enhancement of spin correlations along the $J_1$ chains is accompanied by the weakening in the perpendicular direction. In contrast, the correlations between the chains become stronger for $J_1'/J_1>1$. Although such a behavior is not surprising and follows the simple physical picture of strong and weak bonds, the similarity of the curves corresponding to different $J_1'/J_1$ is remarkable. Fig.~\ref{cu2v2o7_fig_correl} evidences that the spin correlations strongly resemble the behavior of the isotropic honeycomb lattice, even for a sizable spatial anisotropy $J_1'/J_1$. This suggests that despite the spatial anisotropy,
$\beta$-Cu$_2$V$_2$O$_7$ is a surprisingly good realization of the spin-1/2 Heisenberg model on the honeycomb lattice.
\section{Discussion and conclusions}
\label{discussion}
Based on a thorough microscopic study, we interpret the magnetic behavior of $\beta$-Cu$_2$V$_2$O$_7$ within the spatially anisotropic honeycomb model. Recent theoretical reports show that the properties of this model depend on the magnitude of the spatial anisotropy.\cite{li2010} For strong dimer anisotropy ($J_1'/J_1>1.85$), a spin gap is opened, typical for a system of weakly coupled spin dimers. In contrast, a weak anisotropy leads to the honeycomb-lattice physics with the N\'eel AFM ordering. Our estimates of individual exchange couplings suggest $J_1'/J_1=0.5-1.3$ and place $\beta$-Cu$_2$V$_2$O$_7$ in the region of the N\'eel ordering, consistent with the experiment.\cite{ueda2008} An AFM coupling $J_d$ will further stabilize this ordering, while a FM $J_d$ will induce magnetic frustration. Our estimate of $|J_d|/\bar J_1\leq 0.2$ suggests that the effect of $J_d$ should be relatively weak. Nevertheless, theoretical studies of the $J_1-J_1'-J_d$ model with the non-obvious \emph{ferromagnetic} third-neighbor coupling $J_d$ are desirable and could be relevant for related systems.

The magnetic structure of $\beta$-Cu$_2$V$_2$O$_7$ should bear further signatures of the underlying honeycomb spin lattice. The pronounced two-dimensionality and the low coordination number (three bonds per site) will lead to a strongly suppressed ordered moment of about $0.45-0.55$~$\mu_B$,\cite{loew2009} well below the classical value. A similar ordered moment is found in green dioptase with its peculiar 3D spin lattice that comprises three bonds per site only.\cite{dioptase} In $\beta$-Cu$_2$V$_2$O$_7$, the spin arrangement should feature antiparallel ordering along $J_1$ and $J_1'$ that further leads to an antiparallel ordering within the structural dimers (see Fig.~\ref{honeycomb}). The expected propagation vector is $\mathbf k=0$ (with respect to the atomic lattice), but the $C$-centering of the atomic structure has to be broken. In contrast, the AFM ordering within the $J_a-J_d$ chains does not violate the $C$-centering. Thus, neutron diffraction experiments are a feasible test for the proposed 2D model of $\beta$-Cu$_2$V$_2$O$_7$. Inelastic neutron scattering should be able to resolve individual couplings and to give a direct confirmation for the leading couplings $J_1$ and $J_1'$. Additionally, the excitation spectrum of the spin-$\frac12$ honeycomb lattice can be studied.

$\beta$-Cu$_2$V$_2$O$_7$ gives an instructive example of the non-trivial magnetic interactions in transition-metal compounds. The symmetry of the magnetic orbital along with the Cu--O--Cu angle of $98.7^{\circ}$ in the structural dimer disfavor sizable exchange couplings between the structural nearest neighbors. Then, the long-range couplings come into play. The situation largely resembles VO(HPO$_4)\cdot 0.5$H$_2$O (Ref.~\onlinecite{tennant1997}) or Cu$_2$(PO$_3)_2$CH$_2$ (Ref.~\onlinecite{schmitt2010}) where spin dimers do not coincide with the structural dimers. Another relevant example is BiCu$_2$PO$_6$ with its rungs of the spin ladder running \emph{between} the structural ribbons.\cite{mentre2009,bicu2po6} Such intricate implementation of the long-range superexchange couplings is an abundant well of surprises, as in the present study. $\beta$-Cu$_2$V$_2$O$_7$ reveals a spin-$\frac12$ honeycomb lattice, despite lacking any apparent structural features of the honeycomb geometry. We note that the isostructural Cu$_2$P$_2$O$_7$ and Cu$_2$As$_2$O$_7$ compounds could show similar features and deserve further investigation.

In summary, we have shown that $\beta$-Cu$_2$V$_2$O$_7$, previously considered as a spin-chain compound, should be consistently described as a honeycomb-lattice system. The leading couplings $J_1$ and $J_1'$ run via the non-magnetic VO$_4$ tetrahedra, while the couplings between the structural nearest neighbors are weak. The averaged coupling $\bar J_1$ amounts to $60-66$~K. The spatial anisotropy is relatively small and has little influence on the ordered moment as well as on spin correlations. The interlayer coupling $J_{\perp}\simeq 13$~K leads to the N\'eel antiferromagnetic ordering at $T_N=25-30$~K. We propose that $\beta$-Cu$_2$V$_2$O$_7$ shows a relatively low sublattice magnetization of $0.45-0.55$~$\mu_B$, typical for the spin-$\frac12$ honeycomb lattice. Further tests of our model should include high-field magnetization measurements, neutron diffraction, and inelastic neutron scattering. We conclude that $\beta$-Cu$_2$V$_2$O$_7$ is a noteworthy experimental realization of the spin-$\frac12$ Heisenberg model on the honeycomb lattice. The convenient energy scale and the lack of disorder make it a promising system to challenge experiment and theory and to improve our understanding of low-dimensional magnets.

\acknowledgments
We acknowledge Vladimir Mazurenko and Vladimir Anisimov for providing the TB-LMTO-ASA code with the LEIP option. We are also grateful to Johannes Richter, Alim Ormeci, and Juri Grin for careful reading of the manuscript and fruitful discussions. A.T. acknowledges the financial support from Alexander von Humboldt Foundation.


\begin{thebibliography}{67}
\bibitem [{not(){\natexlab{b}}}]{note2}%
  \BibitemOpen
  \bibinfo {note} {(VO)$_2$P$_2$O$_7$ has several polymorphic modifications
  with similar structures. $\beta$-Cu$_2$V$_2$O$_7$ is closest to the
  high-pressure form of (VO)$_2$P$_2$O$_7$.}\BibitemShut {Stop}%
\end{thebibliography}

%

\end{document}